\documentclass[11pt,twoside]{article}
\usepackage[dvips]{graphics}
\usepackage{newpasp}
\newcommand{\oversim}[2]{\protect{\mbox{\lower0.5ex\vbox{%
   \baselineskip=0pt\lineskip=0.2ex
   \ialign{$\mathsurround=0pt #1\hfil##\hfil$\crcr#2\crcr\sim\crcr}}}}} 
\newcommand{\simgreat}{\mbox{$\,\mathrel{\mathpalette\oversim>}\,$}} 
\newcommand{\simless} {\mbox{$\,\mathrel{\mathpalette\oversim<}\,$}} 
\def\edcomment#1{\iffalse\marginpar{\raggedright\sl#1\/}\else\relax\fi}
\reversemarginpar

\begin{document}
\title{Stellar-Dynamics of Young Star Clusters}
 \author{Pavel Kroupa\footnotemark[1]} \footnotetext[1]{On
 leave from the Institut f\"ur Theoretische Astrophysik,
 Tiergartenstr. 15, D-69121 Heidelberg (postal address)}

\affil{Max-Planck-Institut f\"ur Astronomie, Heidelberg, Germany}

\begin{abstract}
The stellar-dynamical evolution of bound star clusters during the
first few~Myr is dominated by binary-binary and binary-star
interactions, the rapid sinking of the most massive stars to the
centre of the clusters and mass loss from evolving stars. The
consequences of these processes for the binary and stellar population
in clusters, and for the star clusters as a whole, are studied by
following the evolution over 150~Myr of a library of compact cluster
models containing up to $10^4$~stars.
\end{abstract}


\section{Introduction}
The evolution of the primordial binary population, mass segregation,
the ejection of massive stars, and the effect of mass loss through
evolving stars, are dynamical processes that shape the stellar content
and structure of young clusters.  A particularly interesting problem
is the attempt to reconstruct the initial configuration of young
clusters, such as the Orion Nebula Cluster and the Pleiades, by using
the binary population as a memory agent for the dynamical history of a
population.  Since most stars probably form in clusters of some sort,
the multiple-star properties of the stellar population in the Galactic
field is expected to be significantly shaped by the dynamical
processes in these.

To understand the appearance of star clusters of any age, their
dynamical and astrophysical evolution needs to be treated accurately
and consistently.  Direct $N$-body integration with special
mathematical methods to treat accurately and efficiently multiple
close stellar encounters as well as perturbed binary and higher-order
systems, together with modern stellar evolution algorithms, is
achieved with the {\sc Nbody6} programme (Aarseth 1999).  Using a
specially modified version of this code, the evolution of clusters
with $N\le10^4$ stars is presented here.

\section{Models}
Table~\ref{tab:mods} lists the cluster models.  Each was followed for
150~Myr.  $N$ is the total number of stars, and $N_{\rm bin}$ the
number of primordial binaries.  Birth binary proportions $f_{\rm b}=1,
0.6, 0$ are used with distributions of binding energies as in
Taurus--Auriga or the Galactic field (Kroupa, Petr, \& McCaughrean
1999, hereinafter KPM), respectively. In all cases the central number
density has the value observed in the Trapezium Cluster, $\rho_{\rm
c}=10^{4.78}$~stars/pc$^3$, and the half-mass-\-diameter crossing time
$t_{\rm cr}=0.24$~Myr. The half-mass radius and three-dimensional
velocity dispersion are $R_{0.5}$ and $\sigma_{\rm 3D}$, respectively;
the initial density distributions being Plummer laws.  All results
shown here are averages of $N_{\rm run}$ calculations per model.  In
all cases the KTG93 IMF (Kroupa, Tout, \& Gilmore 1993) is used for
$0.08-1.0\,M_\odot$. For all but the 100f10 models, a power-law IMF
with $\alpha=2.3$ (Salpeter: $\alpha=2.35$) for $m>1\,M_\odot$, and
with $\alpha=0.3$ for $0.01\le m<0.08\,M_\odot$, is employed. The
100f10 models have $\alpha=2.7$ for $m>1\,M_\odot$, and $\alpha=0.5$
for brown dwarfs.  The mass range is thus $0.01-50\,M_\odot$ giving
the average stellar mass $<\!\!m\!\!>$.  Random pairing from the IMF
is assumed to give the primordial mass-ratio distribution.  The
resulting median relaxation time, $t_{\rm rel}$, assumes the cluster
either consists of centre-of mass systems (short values) or only
single stars. Owing to the rapidly changing {\it number of systems}
the true $t_{\rm rel}$ lies between these two values.

\vskip -5mm
\begin{table}
\caption{Cluster models}
\vskip 1mm
\begin{tabular}{cccccccc}
\tableline
model &$N$ &$N_{\rm bin}$ &$R_{0.5}$ &$<\!\!m\!\!>$ 
&$\sigma_{\rm 3D}$ &$t_{\rm rel}$ &$N_{\rm run}$\cr

      &    &              &[pc]      &[$M_\odot$]           
&[km/s]            &[Myr] \cr
\tableline
8f10   &800   &400    &0.19    &0.37 &1.6 &0.8--1.4 &10 \cr
8f06   &800   &300    &0.19    &0.37 &1.6 &1.0      &10 \cr
8f00   &800   &0      &0.19    &0.37 &1.6 &1.4      &5  \cr

30f10  &3000  &1500   &0.30    &0.37 &2.5 &2.4--4.4 &3 \cr
30f06  &3000  &1125   &0.30    &0.37 &2.5 &3.0      &3 \cr
30f00  &3000  &0      &0.30    &0.37 &2.5 &4.4      &3 \cr

100f10 &$10^4$ &5000 &0.45     &0.28 &3.2 &14.5     &2 \cr
\tableline
\tableline
\end{tabular}\label{tab:mods}
\end{table}
\vskip -3mm

\section{Stimulated binary evolution}
Binary systems are ubiquitous in all known stellar populations. In
particular, open clusters such as the Pleiades may contain a binary
proportion up to 70~per cent ($f_{\rm tot}=0.70$) (K\"ahler 1999).  In
the Galactic field, $f_{\rm tot}\approx0.6$, with a Gaussian
distribution of log$_{10}P$, where $P$ is the binary-star period in
days. In sparse nearby star-forming regions $f_{\rm b}\approx1$, with
a primordial period distribution that is consistent with rising to the
longest periods, and $f_{\rm tot}\simgreat0.5$ even in the highly
concentrated very young Trapezium Cluster (see KPM and references
therein).  Realistic star-cluster models should therefore contain
$f_{\rm b}>0.5$ at birth.

\begin{figure}
\plotfiddle{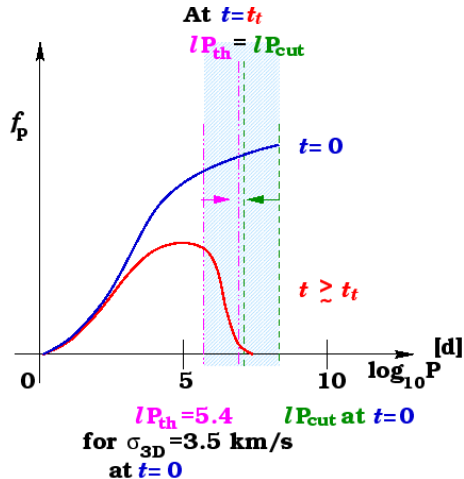}{6cm}{0}{40}{40}{-130}{-70}
\caption{Illustration of the stimulated evolution of the distribution
of binary star periods in a cluster ($l{\rm P}={\rm log}_{10}P$). A
binary has orbital period $P_{\rm th}$ when $\sigma_{\rm 3D}$ equals
its circular orbital velocity.}
\label{fig:stev}
\end{figure}

The evolution of binary orbital parameters through encounters with
other systems is termed {\it stimulated evolution}.  During the
initial few crossing times the weakly bound binary systems are
disrupted which can cool the cluster (KPM). This occurs for those
binaries that have a binding energy, $|e_{\rm b}|$, smaller than the
average kinetic energy of the cluster stars, ${\overline e}_{\rm k}$.
Binaries with $|e_{\rm b}|>{\overline e}_{\rm k}$, on the other hand,
are {\it hard}, and on average they harden further through
interactions with cluster stars thereby heating the cluster (Heggie
1975).  At any time, binaries near the {\it hard/soft} boundary, with
periods $P\approx P_{\rm th}$ denoting the {\it thermal period}, are
most active in the energy exchange between the cluster field and the
binary population.  The cluster expands as a result of binary heating
and mass segregation (see below), and the hard/soft boundary, $P_{\rm
th}$, shifts to longer periods (Fig.~\ref{fig:stev}).  Meanwhile,
binaries with $P>P_{\rm th}$ continue to be disrupted while $P_{\rm
th}$ keeps shifting to longer periods. This process ends when $P_{\rm
th}\simgreat P_{\rm cut}$, which is the cutoff or maximum period in
the surviving period distribution. At this critical time, $t_{\rm t}$,
further cluster expansion can be reduced since the population of
heating sources, i.e. the binaries with $P\approx P_{\rm th}$, is
significantly reduced. The details strongly depend on the initial
value of $P_{\rm th}$ by determining the amount of binding energy in
soft binaries which can cool the cluster if significant enough.

Stimulated evolution also leads to a depletion of the {\it mass-ratio
distribution} at small values, but is not efficient enough to
thermalize the {\it eccentricity distribution}, requiring binaries to
form with the observed thermal eccentricity distribution (Kroupa 1995).

After the critical time, $t_{\rm t}$, the expanded cluster has reached
a temporary state of thermal equilibrium, and further stimulated
evolution will occur with a significantly reduced rate determined by
the velocity dispersion in the cluster, the cross section given by the
semi-major axis of the binaries, and their number density and that of
single stars in the cluster.  Stimulated evolution during this slow
phase will usually involve partner exchanges and unstable but also
long-lived {\it hierarchical systems}. The {\it IMF} is critically
important for this stage, as the initial number of massive stars
determines the cluster density at $t\simgreat5$~Myr. Further binary
depletion will occur once the cluster goes into core-collapse and the
kinetic energy in the core rises. This phase of advanced cluster
evolution is, however, not addressed here.

\begin{figure}
\plotfiddle{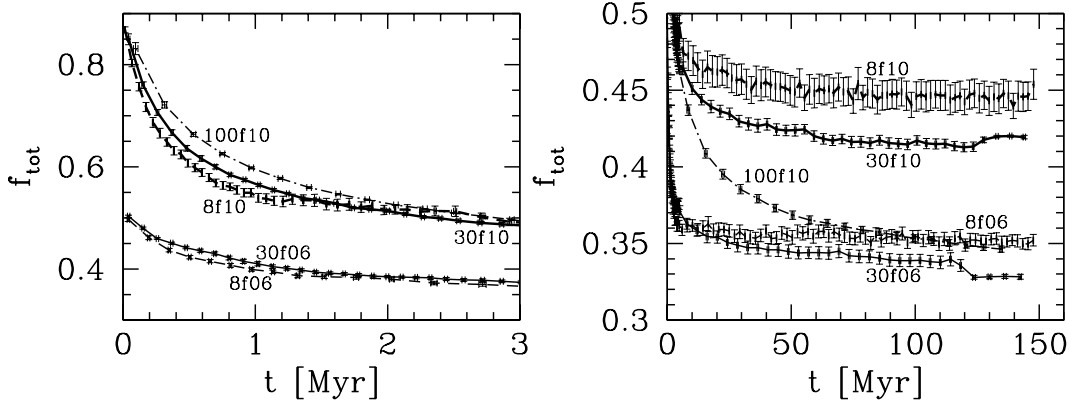}{4.2cm}{-90}{55}{55}{-210}{315}
\caption{Evolution of the total binary population for models 8f10
(thick dashed line), 8f06 (medium dashed line), 30f10 (thick solid
line), 30f06 (medium solid line) and 100f10 (dash-dotted line).}
\label{fig:ftot}
\end{figure}

Fig.~\ref{fig:ftot} illustrates the evolution of the binary
proportion. All systems are counted within and outside the clusters.
Initially, the binary proportion decays on the crossing time-scale,
but slows significantly in all cases after $t\approx1$~Myr
$\approx4\,t_{\rm cr}\approx t_{\rm t}$. {\it The value of $f_{\rm b}$
is remembered by 150~Myr for the same IMF}. However, for model~100f10
$f_{\rm tot}$ crosses over to the $f_{\rm b}=0.6$ cases (models~8f06
and 30f06) which is due to the smaller number of massive stars.
Cluster expansion owing to stellar evolutionary mass loss is lessened
in model~100f10, and consequently, the clusters retain a higher
density.

Stimulated evolution also leads to the generation of binary systems
with {\it forbidden orbits}. These binaries have an eccentricity, $e$,
that places them above or to the left of the upper envelope in $e-{\rm
log}_{10}P$ diagrams evident in binary populations (Kroupa
1995). Forbidden orbits are expected to circularise (i.e. {\it
eigenevolve}) rapidly. This can lead to coalescence in some cases, but
when $e\approx1$ after an encounter, the two stars can collide and
merge within an orbital period. Since the cross section for encounters
is much larger for binaries than for single stars, this channel
leading to {\it stellar collisions} will dominate throughout the
cluster life-time, possibly leading to exotic very massive stars in
the early phase and/or blue stragglers in older clusters. This is also
discussed by Portegies-Zwart (2000) for clusters with extraordinarily
high central densities and assuming stars stick once they touch.

\begin{figure}
\plotfiddle{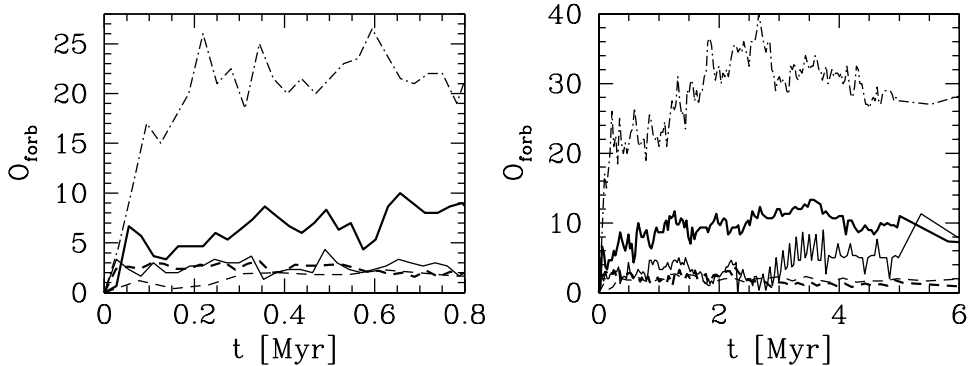}{4cm}{-90}{50}{50}{-200}{285}
\caption{The number of forbidden orbits at a given time. Models are as
in Fig.~\ref{fig:ftot}.}
\label{fig:nf}
\end{figure}

A more conservative approach is taken here by counting the number of
forbidden orbits, $O_{\rm forb}$, present at discrete time-intervals
(Fig.~\ref{fig:nf}).
The figure shows that larger $f_{\rm b}$ leads to larger $O_{\rm
forb}$, and that $O_{\rm forb}$ increases significantly during
$t\simless t_{\rm cr}$. After this time, the generation of further
forbidden orbits is reduced, and $O_{\rm forb}$ is about 0.3~\% of the
initial number of systems when $f_{\rm b}=0.6$, but about 0.7~\%
when $f_{\rm b}=1$. The maximum in $O_{\rm forb}$ for $t\simgreat
2$~Myr when $N\ge3000$ is interesting, but remains to be investigated.

\section{Cluster evolution}
\begin{figure}
\plotfiddle{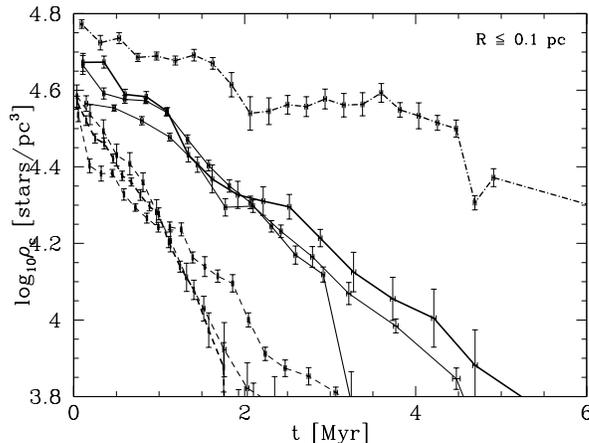}{5cm}{-90}{30}{30}{-120}{165}
\caption{Evolution of the central number density. Models are as in
Fig.~\ref{fig:ftot}. In addition, single-star models 8f00 (thin dashed
line) and 30f00 (thin solid line) are shown.}
\label{fig:cd}
\end{figure}

Three mechanisms drive internal cluster evolution: (i) energy
equipartition (two-body relaxation), (ii) binary star activity, and
(iii) stellar evolution.  Fig.~\ref{fig:cd} illustrates the evolution
of the central number density, which is defined by the number of stars
within the central radius of 0.1~pc.
The central number density decreases on the relaxation time-scale. It
also decreases for the clusters with $f_{\rm b}=0$. This is due to the
shedding of binding energy, gained by the massive stars as they
rapidly sink to the cluster centre as a result of energy
equipartition, to the cluster field. The central average mass
increases by a factor of~2 within~1~Myr, and a factor of~4 by~2~Myr,
in clusters with the Salpeter IMF above $1\,M_\odot$, independently of
the binary proportion. The cluster (100f10) with the steeper IMF for
massive stars also shows a linear increase of the central average
mass, but the rate of increase is a factor of~2 slower.

\begin{figure}
\plotfiddle{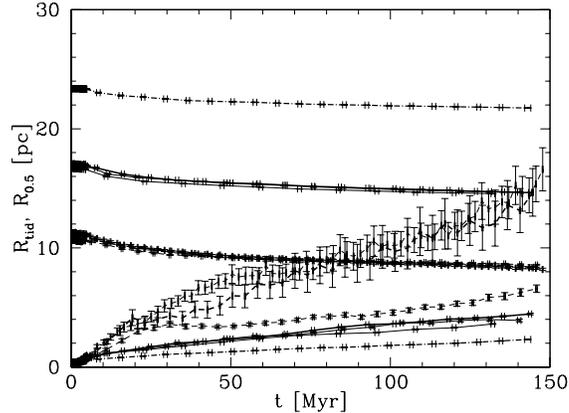}{4.5cm}{-90}{28}{28}{-110}{155}
\caption{Tidal and half-mass radii. Models are as in 
Fig.~\ref{fig:cd}.}
\label{fig:rtrh}
\end{figure}

Returning to the central number density, the clusters with $f_{\rm b}>0.5$
expand at a faster rate than the single-star clusters, which is due to
heating from the significant binary population.  The higher density in
model~30f10 (thick solid line) than in model~30f06 (medium solid line)
for $t\simless0.6$~Myr is significant and stems from the {\it cooling}
of the cluster through disrupting wide binaries in~30f10. This effect
is seen in reverse in models~8f$x$ (thick and medium dashed lines)
because the initially smaller $\sigma_{\rm 3D}$ makes more of the
binary population effectively hard, thus leading to more heating in
the case $f_{\rm b}=1$.  The overall bulk evolution is illustrated in
Fig.~\ref{fig:rtrh}.
The tidal radius, $R_{\rm tid}$, decreases as stars escape, and
$R_{0.5}$ increases owing to cluster expansion. The clusters composed
of single stars only (8f00, 30f00) expand at a slower rate because
binary-heating is not effective. Only the models with $N=800$ stars
loose approximately 50~\% of their mass within about 150~Myr.

\begin{figure}
\plotfiddle{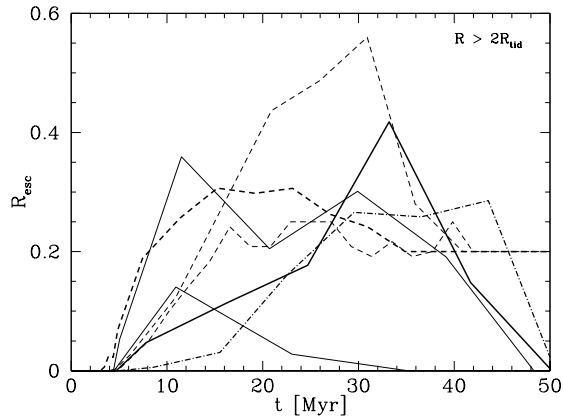}{4.5cm}{-90}{28}{28}{-110}{155}
\caption{The relative number of massive stars ($m\ge8\,M_\odot$) 
at distances larger than $2\,R_{\rm tid}$. Models are as in 
Fig.~\ref{fig:cd}.}
\label{fig:nesc}
\end{figure}

Owing to the rapid onset of {\it mass segregation}, stars and systems
with lower mass have, on average, larger radii than those with higher
mass. In particular, the calculations show that {\it brown dwarfs}
have the largest mean radius and are preferentially lost from the
cluster. However, not only low-mass stars escape. The massive stars
sink to the cluster core on a time-scale given roughly by
$\left(<\!\!m\!\!>/m_{\rm mass}\right)t_{\rm rel}$ which is
$\simless0.1 t_{\rm rel}$ for the cases studied here. The massive
stars interact energetically with each other, often leading to their
ejection. Many calculations show two~O and/or~B stars leaving a
cluster in opposite directions. This occurs if at least one of the
stars is a binary system that gains binding energy and propels itself
and the other star out of the cluster. Fig.~\ref{fig:nesc} shows the
ratio $R_{\rm esc}=N_{\rm esc}/N_{\rm now}$, where $N_{\rm esc}$ is
the number of stars with $m\ge8\,M_\odot$ and $R>2\,R_{\rm tid}$, and
$N_{\rm now}$ is the number of such stars remaining in the calculation
($R_{\rm esc}=0$ if $N_{\rm now}=0$).
Thus, between about~5 and 50~Myr, 10--50~\% of all massive stars are
found at distances larger than $2\,R_{\rm tid}$
(Fig.~\ref{fig:rtrh}). This has significant implications for studies
of the IMF in young clusters, as well as in the field. The figure also
shows that there is no significant difference between the cluster
models. However, more massive clusters have a larger $R_{\rm tid}$, so
that it takes longer for stars with equal velocity to reach $2\,R_{\rm
tid}$ for massive clusters.

\section{Conclusions}
Binary--binary and binary--single-star interactions dominate the
dynamical evolution during the first few crossing times, during which
time the population of soft binaries is depleted significantly, and
binaries near the hard/soft boundary contribute to cluster heating.
The rapid sinking of the most massive stars to the centre of the
cluster, on a time-scale which is a small fraction of the relaxation
time, also leads to cluster heating and consequently cluster
expansion.  About~0.3 and 0.7~\% of the initial number of systems for
$f_{\rm b}=0.6$ and~1, respectively, are hardened sufficiently during
an encounter such that efficient circularisation sets in. A fraction
of such binaries will have their constituent stars collide.  Massive
stars are readily ejected from the cluster cores, and between~10 and
50~\% of~O and~B stars are located further than $2\,R_{\rm tid}$ from
their clusters between about~5 and~50~Myr.  The redistribution of
binding and kinetic energy leads to cluster expansion, and the central
density falls on a relaxation time-scale.  After about 4~Myr the
massive stars that remain in the cluster start evolving significantly,
and their combined mass loss leads to further cluster expansion which
increases the evolution time-scale of the cluster.  The number of
massive stars, and thus the IMF, is critical in determining the
expansion of the cluster during this phase, and by how much the
primordial binary population is eroded during the first 150~Myr.  The
proportion of primordial binary systems that survive the first
150~Myr, in clusters that initially have a central number density such
as in the Orion Nebula Cluster, is 34~and 43~\% for $f_{\rm b}=0.6$
and~1, respectively, for a Salpeter IMF ($\alpha=2.3$) above
$1\,M_\odot$.  If $\alpha=2.7$ for $m>1\,M_\odot$, then $f_{\rm
tot}\approx0.35$ despite $f_{\rm b}=1$.  Finally, the brown dwarf
population lies at a larger average radius than more massive stars
during cluster evolution.

The results presented here are preliminary in that a larger model
library is being completed now, and the formidable data reduction is
being advanced. Many questions remain to be answered.  In particular,
the effect of gas ejection from a newly fledged cluster needs to be
addressed. Work on this and other issues is in progress.

\vskip 5mm
\noindent Acknowledgements
\vskip 2mm 
\noindent
It is a pleasure to thank Sverre Aarseth and Jarrod Hurley for making
{\sc Nbody6} freely available and aiding in code development specific
for this project, which is supported through DFG grant KR1635.

\section{Discussion}
{\bf David Schaerer:} 1) Is there a contradiction between your
scenario where massive stars are very rapidly ejected from clusters
and the scenario of Bonnell, Davies, \& Zinnecker (1998) requiring
massive stars to form in dense clusters?

\noindent {\bf PK:} No, not at all. It supports it since the location
of massive stars away from clusters can be interpreted readily as being
due to dynamical ejection. Velocity measurements, however, are
required to confirm this in any particular case. 

\noindent 2) Could you comment on the implications of your scenario on
the IMF of massive stars? Also, Massey finds indications for an IMF
{\it less} populated by massive stars in the field compared to
clusters. How does this fit into your picture? 

\noindent {\bf PK:} This is a very important question which I am
working on. Indeed, the MF for massive stars measured in clusters
cannot be taken to be the IMF (even if stellar evolution is taken
account of), and the field stars are likely to have been ejected from
clusters. Qualitatively this agrees with the finding by Massey that
the field IMF is steep, because less massive OB stars have a higher
chance of being ejected with large velocities than the massive
ones. However, this requires extensive numerical experiments to
quantify the ejection velocities and ejected mass spectrum, before it
can be ascertained that a ``primordial field'' population of OB stars
with its own IMF does not exist.

\vskip 3mm

\noindent
{\bf Hans Zinnecker:} I noticed that the time scale for mass
segregation you gave for the Orion Nebula Cluster was shorter than the
crossing time. How can that be? If mass segregation is so fast, does
that mean that the central Trapezium stars may not have been born in
the very centre? In other words: do you disagree with the recent
simulations of Bonnell \& Davies (1998)?

\noindent {\bf PK:} The timescale for mass segregation is not well
defined. Spitzer estimated the timescale for energy equipartition
assuming an isothermal stellar population. This estimate is $t_{\rm
equ}=\left(<\!\!m\!\!>/m_{\rm mass}\right)t_{\rm rel}$ which can be,
formally, shorter than $t_{\rm cross}$. However, $t_{\rm equ}$ is only
the timescale for the {\it onset} of mass segregation, and as soon as
the density distribution has changed as a result of energy
equipartition, the estimate is no longer applicable. Only numerical
experiments can lead to a better understanding of the mass-segregation
time-scale. But little work has been done so far on
this. Unfortunately Bonnell \& Davies used a code that was developed
by Aarseth for galaxy simulations, and they do not specify the value
of the softening parameter they adopted. Therefore, it can, at this
stage, not be said that their calculations correctly treat the
cumulative effects of many close encounters. One important effect not
considered by Bonnell \& Davies is the very high primordial binary
population. This aids mass segregation because the binaries have a
significantly larger encounter cross section.

\end{document}